\documentclass{article}
\usepackage{amsmath}
\usepackage{amssymb}
\usepackage{makeidx}
\usepackage{latexsym}
\usepackage[dvips]{graphicx}
\usepackage{textcomp}

\begin{document}

\title{{\huge Neutrinos superluminality and\\ Local Lorentz Invariance}}

\author{{ Fabio Cardone}$^{1,2}${ , Roberto Mignani}$^{2,3,4}$%
{ \ and Andrea Petrucci}$^{1,*}$ \\
\\
$^{1}$Istituto per lo Studio dei Materiali Nanostrutturati (ISMN -- CNR)\\
Via dei Taurini - 00185 Roma, Italy\\
$^{2}$GNFM, Istituto Nazionale di Alta Matematica "F.Severi"\\
\ Citt\`{a} Universitaria, P.le A.Moro 2 - 00185 Roma, Italy\\
$^{3}$Dipartimento di Fisica \textquotedblright
E.Amaldi\textquotedblright ,
Universit\`{a} degli Studi \textquotedblright Roma Tre\textquotedblright \\
\ Via della Vasca Navale, 84 - 00146 Roma, Italy\\
$^{4}$\,I.N.F.N. - Sezione di Roma Tre \\ $^{*}$corresponding author
- petruccia@fis.uniroma3.it
+393392573280\\
\\
} \maketitle

\date{}

\begin{abstract}
The recent measurement of the neutrino velocity with the OPERA
detector in the CNGS beam, on whose basis it was found that (v-c)/c
= (2.48 $\pm$ 0.28 (stat.) $\pm$ 0.30 (sys.)) $\cdot$ $10^{-5}$,
does not contain any significant violation of Local Lorentz
Invariance (LLI), since the corresponding value of the parameter
$\delta$=$(u/c)$$^{2}$-1, that represents the upper limit of the
breakdown of LLI, is at least
 three orders of magnitude higher than the known lower limit reported in literature and is compatible with the values estimated
by other experiments carried out so far.

\end{abstract}

\section{Local Lorentz Invariance}

The fundamental teaching of Einstein's relativity theories is that
physical phenomena occur in  a four-dimensions space-time possessing
a global curved (Riemannian) structure and a local flat
(Minkowskian) one. This implies the existence of a local frame in
which Special Relativity (SR) strictly holds for non-gravitational
interactions. Such a property is referred to as local Lorentz
invariance (LLI). However, it is an old-debated problem whether LLI
preserves its validity at any length or energy scale (far enough
from the Planck scale, when quantum fluctuations are expected to
come into play). From the experimental side, the main tests of LLI
which have been carried out up to now can be roughly divided in
three groups:
\begin{enumerate}
\item[(a)]
Michelson--Morley-type (MM) experiments, aimed at testing
isotropy of the round-trip speed of light;
\item[(b)] tests of the isotropy of the one-way speed of light (based on atomic spectroscopy
and atomic timekeeping);
\item[(c)] Hughes--Drever-type (HD)
experiments, testing the isotropy of nuclear energy levels.
\end{enumerate}

\begin{figure}\label{lli break}
\begin{center}
\includegraphics[width=0.8\textwidth]{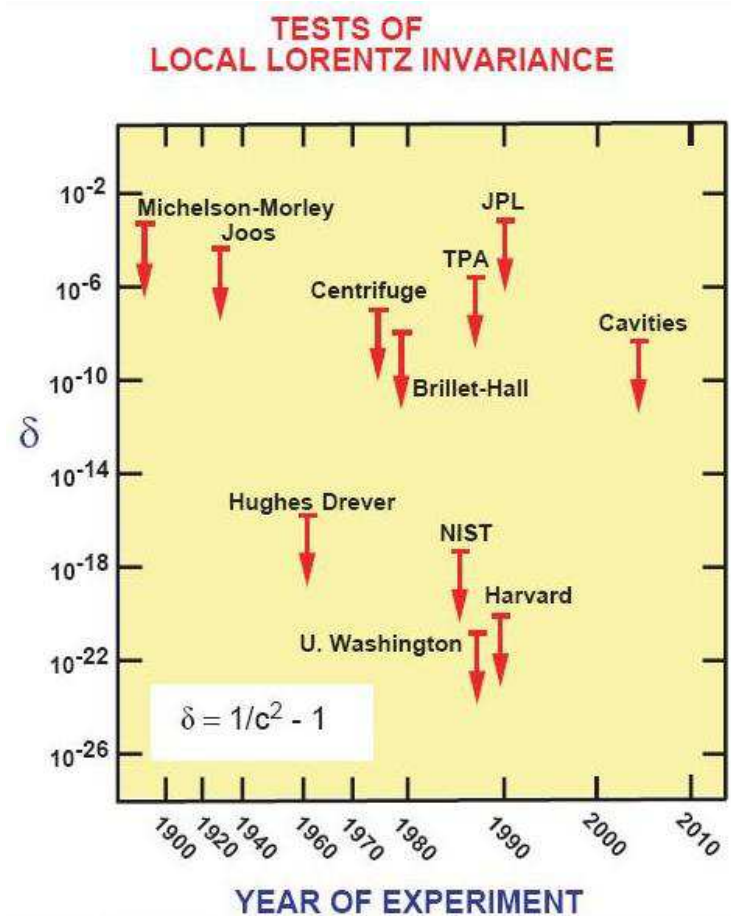}
\caption{\textit{The present experimental situation of the limits of
the LLI breakdown parameter $\delta$ (from \cite{Will_new}, p. 12).
}}
\end{center}
\end{figure}

All such experiments set upper limits on the degree of violation of
LLI. In Fig.\ref{lli break} it is shown the present experimental
situation of the limits of the LLI breakdown parameter\footnote{The
meaning of the parameter [$\delta=((c+v)/c)^{2}-1$] is strictly
connected with the existence of a drift velocity '\emph{v}' (the
velocity of the reference frame where the experiment is performed
with respect to the mean rest frame of the universe or equivalently
to the frame of the CMB). The lower the parameter $\delta$, the
lower this velocity. Many different attempts with increasing
resolution to measure '\emph{v}' have been carried out so far since
the first experiment in 1881 by Michelson. In all of them, the
measured values of '\emph{v}' (or $\delta$) have always been lower
than the expected estimated values (taking as drift velocity the
orbital speed of Earth, or other velocities like that of the
rotation of the galaxy). This is why the measured value of \emph{v}
has always been considered the upper limit of the drift velocity. In
this sense, if it actually existed a drift velocity (and hence a
value of $\delta$) or in other words if LLI were broken, this would
be true for a value of \emph{v} (or $\delta$) lower than the
measured upper limit. The existing upper limits for \emph{v} or
$\delta$ in Fig.\ref{lli break} range from $10^{-21}$ up to
$10^{-4}$. However, they have to be grouped from $10^{-8}$ up to
$10^{-4}$ and from $10^{-21}$ up to $10^{-16}$. The first group
refers to experiments based on the electromagnetic interaction,
while the second are based on the hadronic interaction with some
contribution of the electromagnetic and leptonic ones. Thus,
whenever a value for $\delta$ is obtained from an experiment, it has
to be referred to the right group according to the type of
interaction involved in the experiment and not just referred to
energy level involved.} $\delta=(u/c)^{2}-1$ as it is reported in
\cite{Will,Will_new}. As to the result obtained by the OPERA
detector and the CNGS beam, one might put forward the hypothesis
that it shows a violation of Local Lorentz Invariance. By this
letter we want to clarify that this cannot be the case, since the
value of the parameter $\delta$ for this experiment is perfectly
compatible with other already known values obtained by other
experiments\footnote{One might claim the violation of LLI only if
the value of $\delta$ for the experiment with OPERA were a good bit
lower than the known lowest value.}.

\section{Consistency of OPERA measurements with $\delta$ upper limit}

According to~\cite{opera} the baseline used is~$l=730.085$ km and it
was measured an early arrival time of CNGS muon neutrinos with
respect to the one computed assuming the speed of light in vacuum of
$(60.7 \pm 6.9 (\mbox{stat.}) \pm 7.4 (\mbox{sys.}))$ ns. An
interesting compatible estimated value is reported in
\cite{mign_recam}\footnote{We highlight the remarkable fact that in
\cite{mign_recam} it was estimated a superluminal velocity of the
muon neutrino from its measured negative mass. The corresponding
value of early arrival time is about $10 \cdot 10^{-8}$ seconds
against $6 \cdot 10^{-8}$ seconds measured by OPERA.}. If these
values are used to compute the parameter $\delta$ corresponding to
this experiment, one finds the following. Let us define some
variables and the constant $c$:
\begin{itemize}
\item $c = 299 792 458\, m/s$ (speed of light in vacuum)
\item $l = 730085\, m$
\item $u$: measured superluminal velocity
\item $u = c + v$
\item $T = 2.4 \cdot 10^{-3}$ s (time of flight correponding to $c$)
\item $\tau$: time of flight corresponding to $u$
\item $\tau = T - t$
\item $t = 60.7 \pm 6.9 \mbox{(stat.)} \pm 7.4 \mbox{(sys.)}\, ns$
\end{itemize}

\begin{equation}
\begin{array}{l}
\displaystyle\delta=\left(\frac{u}{c}\right)^2 -1 =
\left(\frac{\frac{l}{T-t}}{\frac{l}{T}}\right)^2-1
    =\left(\frac{T}{T-t}\right)^2-1\\
\displaystyle\phantom{\delta=}
    =\left(\frac{1}{1-\frac{t}{T}}\right)^2-1=
    \left(1+\frac{t}{T}+\left(\frac{t}{T}\right)^2 + \ldots\right)^2 -1
\end{array}
\end{equation}

By neglecting the terms with order greater than one, one gets to

\begin{equation}
\delta=\left(1+\frac{t}{T}\right)^2-1=1+\frac{2t}{T}+\left(\frac{t}{T}\right)^2-1\simeq\frac{2t}{T}
\end{equation}

Thus by substituting the values of the variables reported above, one
obtains the estimation of the parameter $\delta$ for this
experiments:

\begin{equation}
\delta\simeq\frac{2t}{T}=\frac{2 \cdot 60.7 \cdot 10^{-9}}{2.4 \cdot
10^{-3}}=5.06 \cdot 10^{-5}.
\end{equation}

If we refer this value to Fig.\ref{lli break}, we see straightaway
that it is perfectly compatible with the already existing limits for
$\delta$ that lie within [$10^{-8}$ $\div$ $10^{-4}$] obtained by
other experiments\footnote{If one refers to Fig.\ref{lli break}, one
might think that the lowest limit of $\delta$ be $10^{-21}$. However
the group of experiments whose $\delta$ lies within $10^{-21}$
$\div$ $10^{-16}$ are based in their execution on the hadronic
interaction with some contribution of the leptonic and
electromagnetic ones. Conversely, this experiment is a mixture of
leptonic and electromagnetic forces. The bunch of experiments whose
$\delta$ lies within $10^{-8}$ $\div$ $10^{-4}$ refers to
experiments based on electromagnetic interaction and this explains
why we refer to it.}. In this sense, not being a good bit lower than
the known lowest value, it does not add anything new to what we
already know about the upper limit of the violation of
LLI~\cite{Will,Will_new,CASYS09,EG,DST}.

\section{Remarks}

In those experiments, where a time of flight is measured, there
exist two types of superluminality. The first type has to do with
the temporal width of the signal (in our case, the width of the
bunch of neutrinos) and hence with the performance of the detectors.
This first type is not a genuine superluminality. The second type is
related, conversely, to the upper limit of the parameter $\delta$.
In this letter we focused our attention only on the second type.\\
The last remark is about the subtleness that has to be used in
analysing a set of data in order to look for a possible violation of
LLI. It is always necessary to refer the data and the possible
findings about $\delta$ to the interactions involved in the
experiment and not just to the range of energy
involved~\cite{CASYS09,EG,DST,ombra1,ombra2,ombra3,colfi}.

\end{document}